\documentclass[12pt]{article}
\usepackage{amssymb,amsmath,amsthm,framed}

\def\openone{\leavevmode\hbox{\small1\kern-3.3pt\normalsize1}}

\def\biglb{\big[\hspace*{-.7mm}\big[}
\def\bigrb{\big]\hspace*{-.7mm}\big]}

\def\Bigglb{\Bigg[\hspace*{-1.4mm}\Bigg[}
\def\Biggrb{\Bigg]\hspace*{-1.3mm}\Bigg]}
\newtheorem{example}{Example}
\begin{document}

\begin{center}
{\large \bf On the (Non)-Integrability of KdV
Hierarchy  with\\[4pt] Self-consistent Sources}

\bigskip

{\bf V. S. Gerdjikov$^\dag$, G.  G. Grahovski$^{\dag , \ddag}$ and R.  I. Ivanov$^{\ddag}$}

\medskip

{\it $^{\dag}$ Institute for Nuclear Research and Nuclear Energy,\\
Bulgarian Academy of Sciences, 72 Tsarigradsko chaussee, \\
1784 Sofia, BULGARIA }

{\it $^{\ddag}$ School of Mathematical Sciences, Dublin Institute of Technology,\\ Kevin Street, Dublin 8, IRELAND }

\end{center}

\begin{abstract}
\noindent Non-holonomic deformations of integrable equations of
the KdV hierarchy are studied by using the expansions over the
so-called ``squared solutions'' (squared eigenfunctions). Such
deformations are equivalent to  perturbed models with external
(self-consistent) sources. In this regard, the KdV6 equation is
viewed as a special perturbation of KdV equation.  Applying
expansions over the symplectic basis of squared eigenfunctions,
the integrability properties of the KdV hierarchy with generic
self-consistent sources are analyzed.  This allows one to
formulate a set of conditions on the perturbation terms that
preserve the integrability. The perturbation corrections to the
scattering data and to the corresponding action-angle variables
are studied. The analysis shows that although many nontrivial
solutions of KdV equations with generic self-consistent sources
can be obtained by the Inverse Scattering Transform (IST), there
are solutions that, in principle, can not be obtained via IST.
Examples are considered showing the complete integrability of KdV6
with perturbations that preserve the eigenvalues time-independent.
In another type of examples the soliton solutions of the perturbed
equations are presented where the perturbed eigenvalue depends
explicitly on time. Such equations, however in general, are not
completely integrable.
\end{abstract}

\section{Introduction}\label{sec:1}

Nonholonomic deformations of integrable equations attracted the
attention of the scientific community in the last few years. In
\cite{kksst08}, based on the Painlev\'e test, applied to a class
of sixth-order nonlinear wave equations, a list of four equations
that pass the test was obtained. Among the three known ones, there
was a new equation in the list (later known as sixth-order KdV
equation, or just KdV6): \begin{eqnarray}\label{KdV6}
\left(-\frac{1}{4}\partial_x^3 +v_{x}\partial_x +
\frac{1}{2}v_{xx}\right)(v_{t}+v_{xxx}-3v_{x}^2)=0.
\end{eqnarray}
Recognising the potential KdV (PKdV) equation in the second factor
in (\ref{KdV6}) and the Nijenhuis recursion operator for PKdV
$\mathcal{ N}_{\rm PKdV}=(-\frac{1}{4}\partial_x^3
+v_{x}\partial_x + \frac{1}{2}v_{xx})\partial_x^{-1}$, after
change of variables $v=v_{x}$ and $w=v_{t}+v_{xxx}-3 v_{x}^2$, one
can convert (\ref{KdV6}) into a ``potential'' form:
\begin{eqnarray}
&&u_{t}+u_{xxx}-6 u u_{x} -w_{x}=0,\label{GKdV6-1}\\
&&-\frac{1}{4}w_{xxx}+ u w_{x}+ \frac{1}{2} w u_{x}=0,
\label{GKdV6-2}
\end{eqnarray}
or equivalently
\begin{eqnarray}\label{KdV6-1}
\left(-\frac{1}{4}\partial_x^2 +u+
\frac{1}{2}u_{x}\partial^{-1}\right)(u_{t}+u_{xxx}-6uu_{x})=0.
\end{eqnarray}
Here $\partial_x^{-1}$ is a notation for the left-inverse of
$\partial_x$: $\partial_x^{-1}f(x)=\int^xf(y)\, {\rm d}y$. Note
also, that $\partial_x^{-1}$ is a Hamiltonian operator and
$-\frac{1}{4}\partial_x^3 +v_{x}\partial_x + \frac{1}{2}v_{xx}$ is
a symplectic operator. Since both operators
$-\frac{1}{4}\partial_x^3 +v_{x}\partial_x + \frac{1}{2}v_{xx}$
and $\mathcal{ N}_{\rm PKdV}$ are weakly nonlocal, then eqn.
(\ref{KdV6}) possesses the same recursion operator as the PKdV
equation \cite{kksst08,wang09}  and therefore infinitely many
integrals of motion. In the same paper \cite{kksst08}, a Lax pair
and auto-B\"acklund transformations are obtained for KdV6
equation, (\ref{KdV6}). In \cite{rgw08}, applying  Hirota bilinear
method to KdV6 equation, another class (much simpler) of
auto-B\"acklund transformation is obtained.

In \cite{bak08} B. Kupershmidt described (\ref{GKdV6-1}) and
(\ref{GKdV6-2}) as a nonholonomic of the KdV equation, written in
a bi-Hamiltonian form:
\begin{eqnarray}
v_{t}=B^1\left({\delta H_{n+1}\over \delta
v}\right)-B^1(w)=B^2\left({\delta H_{n}\over \delta
v}\right)-B^1(w),\qquad B^2(w)=0\label{GKdV6-BH}
\end{eqnarray}
where
\[
B^1=\partial_x, \qquad  B^2=-\frac{1}{4}\partial_x^3
+v_{x}\partial_x + \frac{1}{2}v_{xx}
\]
are the two standard Hamiltonian operators of the KdV hierarchy,
and
\[
H_1=u, \qquad H_2={u^2\over 2}, \qquad H_3={u^3\over
3}-{u_x^2\over 2}, \dots
\]
are the conserved densities for the same hierarchy. The recursion
operator for the KdV hierarchy is given (after a suitable
rescaling) by $\Lambda=B^2\circ (B^1)^{-1}$. A more general setup
suitable for non-Hamiltonian systems of equations is presented in
\cite{wang09}.

Later on, it was shown in \cite{yz08a,yz08b}  that the KdV6
equation is equivalent to a Rosochatius deformation of the KdV
equation with self-consistent sources. Soliton equations with
self-consistent sources have many physical applications
\cite{vkm90,vkm90b,leon,yz08b}. For example, they describe the
interaction of long and short capillary-gravity waves
\cite{vkm90}. Different classes of exact solutions of (\ref{KdV6})
are obtained in \cite{kksst08,ak08,ksn09,kundu2009,kundu2010}. A
geometric interpretation of the KdV6 equation is given in
\cite{pg09}.

In the present paper, exploiting the potential form of the KdV6
equation (\ref{GKdV6-1}), we study the class of inhomogeneous
equations of KdV type
\begin{eqnarray}\label{eq:KdV-gen}
u_t+u_{xxx}-6uu_x=W_x[u](x),
\end{eqnarray}
with an inhomogeneity/perturbation that presumably belongs to the
same class of functions as the field $u(x)$ (i.e. decreasing  fast
enough, when $|x|\to \infty$).

Generally speaking, the perturbation, as a rule destroys the
integrability of the considered nonlinear evolution equation
(NLEE). The idea of perturbation through nonholonomic deformation,
however, is to perturb an integrable NLEE with a driving force
(deforming function), such that under suitable differential
constraints on the perturbing function(s) the integrability of the
entire system is preserved. In the case of local NLEE's, having a
constraint given through differential relations (not by
evolutionary equations) is equivalent to a nonholonomic
constraint.

To the best of our knowledge, the most natural and efficient way
for studying inhomogeneities/perturbations of NLEE integrable by
the inverse scattering method is by using the expansions over the
so-called ``squared solutions'' (squared eigenfunctions). The
squared eigenfunctions of the spectral problem associated to an
integrable equation represent a complete basis of functions, which
helps to describe the Inverse Scattering Transform (IST) for the
corresponding hierarchy as a Generalized Fourier transform (GFT).
The Fourier modes for the GFT are the Scattering data. Thus all
the fundamental properties of an integrable equation such as the
integrals of motion, the description of the equations of the whole
hierarchy and their Hamiltonian structures can be naturally
expressed making use of the completeness relation for the squared
eigenfunctions and the properties of the corresponding recursion
operator.

This approach was developed first for the Zakharov-Shabat system
\cite{K76,K76a,13,15,KKh81,GeHr1,14}. In particular, in
\cite{GeHr1} a special combination of squared solutions named the
`symplectic  basis' was introduced. Its special property consists
in mapping the variation of the potential of the Zakharov-Shabat
system into the variations of the action-angle variables.

Later such basis was introduced also for the Sturm-Liouville's
operator \cite{KKh81,app,IKK94}. Note that  the scattering data of
the Sturm-Liouville's operator in a generic situation have a pole
for $k=0$ and their variational derivatives with respect to the
potential do not vanish when $|x|\to \infty$. This can lead to to
nontrivial contributions to the associated symplectic and Poisson
structures, coming from the boundary terms \cite{app,fata}. For
more details, we refer to \cite{IKK94,gvy2008} and the references
therein.

The expansions over the squared solutions have been used  for the
study of the perturbations of various completely integrable
systems, see \cite{13,14,15}. As important particular cases these
authors analyzed NLEE with singular dispersion relations and have
shown that whenever the dispersion relation has a pole located on
a given eigenvalue $\lambda$ of the corresponding Lax operator
$L$, then the corresponding evolution is no longer isospectral.
Further generalizations are made in \cite{GI92,g1994}.
 An alternative approach to the same type of equations has been
proposed by Leon in \cite{8,9,10,16}; it is based on the
equivalence of the Zakharov-Shabat system to a $\bar{\partial
}$-problem.


The structure of the present paper is as follows: In Section 2 we
give some background material about the direct scattering problem
for KdV hierarchy. In Section 3 we describe briefly the
generalized Fourier transform for the equations of KdV hierarchy
as a key point for their integrability. This includes the minimal
set of scattering data, their time evolution, the expansions over
the complete set of squared solutions and over the symplectic
basis, action-angle variables, etc. In the next Section 4 we
formulate general conditions on the perturbations (driving forces)
for preserving integrability, treating separately integrable and
non-integrable cases. We provide also a set of nontrivial examples
for both cases. Then, in Section 5 we present examples for
1-soliton solutions of some of the perturbed systems from Section
4.


\section{Preliminaries: Integrability of the Equations of KdV Hierarchy}\label{sec:2}

The spectral problem for the  KdV hierarchy is given by the
Sturm-Liouville equation \cite{ZMNP,IKK94}
\begin{equation}
\label{SPKdV} -\Psi_{xx}+u(x)\Psi=k^2\Psi,
\end{equation}
where $u(x)$ is a real-valued (Schwartz-class) potential  on the
whole axia and $k \in \mathbb{C}$ is spectral parameter. The
continuous spectrum under these conditions corresponds to real
$k$.  We assume that the discrete spectrum consists of finitely
many points $k_{n}=i\kappa _{n}$, $n=1,\ldots,N$ where
$\kappa_{n}$ is real.

The direct scattering problem for (\ref{SPKdV}) is based on the
so-called ``Jost solutions'' $f^+(x,k)$ and
$\bar{f}^+(x,\bar{k})$,given by their asymptotics:
$x\rightarrow\infty$ for all real $k\neq 0$ \cite{ZMNP}:
\begin{equation}
\label{eq6} \lim_{x\to\pm \infty }e^{-ikx} f^\pm(x,k)= 1, \qquad
k\in {\Bbb R}\backslash\{0\}.
\end{equation}
From the reality condition for $u(x)$ it follows that $
 \bar{f}^\pm(x,\bar{k}) = f^\pm(x,-k)$.

On the continuous spectrum of the problem (\ref{SPKdV}) and the
vectors of the two Jost solutions  are linearly related
\footnote{According to the notations used in \cite{ZMNP}
$f^+(x,k)\equiv \bar{\psi}(x,\bar{k})$, $f^-(x,k)\equiv
\varphi(x,k)$.}:
\begin{equation} \label{eq8}
f^{-}(x,k)=a(k)f^+(x,-k)+b(k)f^+(x,k), \qquad
\mathrm{Im}\phantom{*} k=0.
\end{equation}
In addition, for real $k\neq 0$ we have:
\begin{equation}
\label{eq5aa} \bar{f}^{\pm}(x,k)=f^{\pm}(x, -k).
\end{equation}
The quantities $\mathcal{R}^{\pm}(k)=b(\pm k)/a(k)$ are known as
reflection coefficients (to the right with superscript ($+$) and
to the left with superscript ($-$) respectively). It is sufficient
to know $\mathcal{R}^{\pm}(k)$ only on the half line $k>0$.
Furthermore, $\mathcal{R}^{\pm}(k)$ uniquely determines $a(k)$
\cite{ZMNP}.  At the points $\kappa_n$ of the discrete spectrum,
$a(k)$ has simple zeroes i.e.:
\begin{equation}\label{eq:a-n}
    a(k) = (k-i\kappa_n)\dot{a}_n +\frac{1}{2} (k-i\kappa_n)^2\ddot{a}_n
    + \cdots,
\end{equation}
Here, the dot stands for a derivative with respect to $k$ and
$\dot{a}_n\equiv \dot{a}(i\kappa_n)$, $\ddot{a}_n\equiv
\ddot{a}(i\kappa_n)$, etc. At the points of the discrete spectrum
$f^-$ and $f^+$ are again linearly dependent:
\begin{equation} \label{eq200}
f^-(x,i\kappa_n)=b_nf^+(x,i\kappa_n).
\end{equation}
The discrete spectrum is simple, there is only one (real) linearly
independent eigenfunction, corresponding to each eigenvalue
$i\kappa_n$, say $f_n^-(x)\equiv f^-(x,i\kappa_n)$.

The sets
\begin{equation}  \label{eq206} \mathcal{S^{\pm}}\equiv\left \{
\mathcal{R}^{\pm}(k)\quad (k>0),\quad \kappa_n,\quad
R_n^{\pm}\equiv\frac{b_n^{\pm1}}{i\dot{a}_n},\quad n=1,\ldots
N\right\}
\end{equation}
are known as scattering data. Each set -- $\mathcal{S^{+}}$ or
$\mathcal{S^{-}}$ of scattering data uniquely determines the other
one and also the potential $u(x)$ \cite{ZMNP,IKK94,ZF71}.

KdV equation appeared initially as models of the propagation of
two- dimensional shallow water waves over a flat bottom. More
about the physical relevance of the KdV equation can be found e.g.
in \cite{CH93,J02,J03,CL09,I07}.

\section{Generalized Fourier Transforms}\label{sec:3}

A key role in the interpretation of the inverse scattering method
as a generalized Fourier transform plays the so-called
`generating' (recursion) operator: for the KdV hierarchy it has
the form \cite{app}:
\begin{equation}
L_{\pm}=-\frac{1}{4}\partial^2+u(x)-\frac{1}{2}\int_{\pm
\infty}^{x}\mathrm{d}\tilde{x} u'(\tilde{x})\cdot. \label{recurs}
\end{equation}
The eigenfunctions of the recursion operator are the squared
eigenfunctions of the spectral problem (\ref{SPKdV}):
\begin{equation}\label{eq23} F^{\pm}(x,k)\equiv (f^{\pm}(x,k))^2,
\qquad F_n^{\pm}(x)\equiv F(x,i\kappa_n),
\end{equation}
From the completeness of the squared eigenfunctions it  follows
that every function $g(x)$, belonging to the same class of
functions as the potential $u(x)$ of the Lax operator, can be
expanded over the two complete sets of ``squared
solutions''\cite{IKK94}:
\begin{eqnarray}
 g(x)=\pm \frac{1}{2\pi
i}\int_{-\infty}^{\infty}\tilde{g}^{\pm}(k)F^{\pm}_{x}(x,k)\mathrm{d}k\mp\sum_{j=1}^{N}\left(g^{\pm}_{1,j}\dot{F}^{\pm}_{j,x}(x)+
g^{\pm}_{2,j}F^{\pm}_{j,x}(x) \right). \label{1}
\end{eqnarray}

where $\dot{F}^{\pm}_{j}(x)\equiv [\frac{\partial}{\partial k}
F^{\pm} (x,k)]_{k=k_j}$ and the Fourier coefficients are
\begin{eqnarray}
 \tilde{g}^{\pm}(k)&=&\frac{1}{k a^2(k)}\left(g,F^{\mp} \right),
\quad \mathrm{where} \quad \left(g,F\right)\equiv \int_{-\infty}^{\infty} g(x)F(x) \mathrm{d}x, \nonumber \\
 g^{\pm}_{1,j}&=&\frac{1}{k_j \dot{a}_j^2}\left(g,F^{\mp}_j
\right),\qquad g^{\pm}_{2,j}=\frac{1}{k_j
\dot{a}_j^2}\left[\left(g,\dot{F}^{\mp}_j\right)-\left(\frac{1}{k_j}+\frac{\ddot{a}_j}{\dot{a}_j}\right)\left(g,F^{\mp}_j\right)
\right].\nonumber
\end{eqnarray}
Here we assume that in addition $f^+$ and $f^-$ are not linearly
dependent at $x=0$. The details of the derivation can be found
e.g. in \cite{E81,IKK94}.

In particular one can expand the potential $u(x)$, the
coefficients are given through the scattering data
\cite{E81,IKK94}:
\begin{equation}\label{Exp u}
\begin{aligned}
u(x) =\pm \frac{2}{\pi i}\int_{-\infty}^{\infty}
k\mathcal{R}^{\pm}(k)F^{\pm}(x,k)  \mathrm{d}k + {\rm 4i}
\sum_{j=1}^{N} k_j R_j^{\pm}F^{\pm}_{j}(x) ,
\end{aligned}
\end{equation}
\noindent The variation $\delta u(x)$ under the assumption that
the number of the discrete eigenvalues is conserved is
\begin{eqnarray}
  \delta
u(x)=- \frac{1}{\pi }\int_{-\infty}^{\infty}\delta
\mathcal{R}^{\pm}(k)F^{\pm}_x(x,k)\mathrm{d}k\pm 2
\sum_{j=1}^{N}\left[R_j^{\pm}\delta
k_j\dot{F}^{\pm}_{j,x}(x)+\delta R_j^{\pm}F_{j,x}^{\pm}\right].
\label{Exp delta u}
\end{eqnarray}
An important subclass of variations are due to the time-evolution
of $u$, i.e. effectively we consider a one-parametric family of
spectral problems, allowing a dependence of an additional
parameter $t$ (time). Then $\delta u(x,t)=u_t \delta t + Q((\delta
t)^2)$, etc. The equations of the KdV hierarchy can be written as
\begin{eqnarray}u_t+\partial_x
\Omega(L_{\pm})u(x,t)=0, \label{KdVH}
\end{eqnarray}
where $\Omega(k^2)$ is a rational function specifying the
dispersion law of the equation. The substitution of (\ref{Exp
delta u}) and (\ref{Exp u}) in (\ref{KdVH})  gives a system of
trivial linear ordinary differential equations for the scattering
data:
\begin{eqnarray}
\mathcal{R}_t^{\pm}\pm 2ik\Omega(k^2)\mathcal{R}^{\pm}&=&0,\label{ScatData 1} \\
R_{j,t}^{\pm}\pm 2ik_j\Omega(k_j^2)R_j^{\pm}&=&0, \label{ScatData 2} \\
k_{j,t}&=&0.\label{ScatData 3}
\end{eqnarray}
The KdV equation $u_t-6uu_x+u_{xxx}=0 $ can be obtained for
$\Omega(k^2)=-4k^2$.

Once the scattering data are determined from (\ref{ScatData 1}) --
(\ref{ScatData 3}) one can recover the solution from (\ref{Exp
delta u}). Thus the Inverse Scattering Transform can be viewed as
a GFT.

For our purposes, it is more convenient to adopt a special set of
``squared solutions'', called symplectic basis \cite{GeHr1,IKK94}.
It has the property that the expansion coefficients of the
potential $u(x)$ over the symplectic basis are the so-called
action-angle variables for the corresponding NLEE.

For the KdV hierarchy, the symplectic basis is given by:
\begin{eqnarray}
\label{kdv-eq60}
\mathcal{P}(x,k)&=&\mp \left(\mathcal{R}^\pm(k)F^\pm(x,k)-\mathcal{R}^\pm(-k)F^\pm(x,-k)\right),\\
\mathcal{Q}(x,k)&=&\mathcal{R}^-(k)F^-(x,k)+\mathcal{R}^+(k)F^+(x,k), \\
P_n(x)&=&-R_n^\pm F_n^\pm(x), \qquad Q_n(x)=
-\frac{1}{2k_n}\Big(R_n^+\dot{F}_n^-(x)-R_n^-\dot{F}_n^+(x)\Big).
\label{kdv-eq61a}
\end{eqnarray}
Its elements satisfy the following canonical relations:
\begin{align}\label{kdv-eq65}
\biglb \mathcal{P}(k_1),\mathcal{Q}(k_2)\bigrb &= \delta(k_1-k_2),
&
 \biglb \mathcal{P}(k_1),\mathcal{P}(k_2) \bigrb
 &=\biglb \mathcal{Q}(k_1),\mathcal{Q}(k_2)\bigrb =0,  \\
 \biglb P_m,Q_n\bigrb &=\delta_{mn}, &  \biglb P_m,P_n\bigrb &= \biglb Q_m,Q_n \bigrb
=0,
\end{align}
($k_{1}>0, k_{2}>0  $) with respect to the skew-symmetric product
\begin{eqnarray}\biglb f,g \bigrb \equiv
\frac{1}{2}\int_{-\infty}^{\infty}(f(x)g_x(x)-g(x)f_x(x)){\rm
d}x=\int_{-\infty}^{\infty}f(x)g_x(x){\rm d}x,\label{kdv-eq31a}
\end{eqnarray}
The symplectic basis  satisfies the completeness relation
\cite{IKK94}:
\begin{multline}\label{kdv-eq63a}
{\theta(x-y)-\theta(y-x)\over 2} =
\frac{1}{2\pi}\int_{0}^{\infty}\Big(\mathcal{P}(x,k)\mathcal{Q}(y,k)-
\mathcal{Q}(x,k),\mathcal{P}(y,k)\Big)\frac{{\rm d}k}{\beta(k)}  \\
- \sum_{n=1}^{N}\Big(P_n(x)Q_n(y)-Q_n(x)P_n(y)\Big),
\end{multline}
where $\beta(k)=2ikb(k)b(-k)$. Notice that the integration over
$k$ is from $0$ to $\infty$. It follows that every function $X(x)$
from the same class as the potential $u(x)$ (i.e. smooth and
vanishing fast enough when $x\to\pm\infty$) can be expanded over
the symplectic basis:
\begin{multline}\label{eq:kdv-X-PQ}
    X(x)= \frac{1}{2\pi}\int_0^\infty \frac{{\rm d}k}{\beta(k)}\, \left( \mathcal{P}(x,k)
    \phi_X(k) - \mathcal{Q}(x,k)\rho_X(k)\right)  \\
    - \sum_{n=1}^N    \left( P_n(x)\phi_{n,X} - Q_n(x)\rho_{n,X}\right).
\end{multline}
The expansion coefficients can be recovered from the so-called
inversion formulas:
\begin{equation}\label{eq:kdv-X-phi}\begin{aligned}
\phi_X(k)  &= \biglb \mathcal{Q}(y,k),X(y)\bigrb , &\qquad
\rho_X(k) &=\biglb \mathcal{P}(y,k),X(y)\bigrb ,  \\
\phi_{n,X} &= \biglb Q_n(y),X(y)\bigrb , &\qquad \rho_{n,X}
&=\biglb P_n(y),X(y)\bigrb .
\end{aligned}\end{equation}
 In particular, if $X(x)=u(x)$ is a solution of the spectral
problem, one can compute \cite{IKK94}:
\begin{align}\label{eq:kdv-u-PQ}
\biglb \mathcal{P}(y,k),u(y)\bigrb &=0, &  \biglb
    \mathcal{Q}(y,k),u(y)\bigrb &=-4ik\beta(k), \\
\biglb P_n(y),u(y)\bigrb&=0, &  \biglb Q_n(y),u(y)\bigrb &=4ik_n .
\label{eq:kdv-u-PQ2}
\end{align}
Thus, from (\ref{eq:kdv-u-PQ}) one gets:
\begin{equation}\label{kdv-eq:q-PQ}
u(x)= \frac{2}{\pi i}\int_0^\infty \mathcal{P}(x,k){\rm d}k
     - \sum_{n=1}^N 4ik_n P_n(x).
\end{equation}
The expression for the variation of the potential is
\begin{multline}\label{eq:kdv-du-PQ}
    \delta u(x)= \frac{1}{2\pi}\int_0^\infty \frac{{\rm d}k}{\beta(k)}\, \left( \mathcal{P}_x(x,k)
    \delta \phi(k) - \mathcal{Q}_x(x,k)\delta \rho(k)\right)  \\
    - \sum_{n=1}^N
    \left( P_{n,x}\delta \phi_{n} - Q_{n,x}\delta \rho_{n}\right)
\end{multline}
with expansion coefficients
\begin{eqnarray} \rho(k)&\equiv &-2ik\ln |a(k)|=-2ik
\ln(1-\mathcal{R}^{-}(k)\mathcal{R}^{-}(-k)), \quad k>0, \label{kdv-eq52} \\
\phi (k)&\equiv &2i\beta(k)\arg b(k)=\beta(k)\ln
\frac{\mathcal{R}^-(k)}{\mathcal{R}^+(k)}, \label{kdv-eq53} \\
\rho_n&=&-\lambda_n=-k_n^2, \qquad \phi_n=2\ln b_n=\ln
\frac{R_n^-}{R_n^+}. \label{kdv-eq54}
\end{eqnarray}
These are known as action ($\rho(k)$) - angle ($\phi (k)$)
variables for KdV equation \cite{ZF71,ZMNP}. They satisfy the
canonical relations
\begin{align}\label{kdv-eq58}
  \{\rho(k_1),\phi(k_2)\}&= \delta(k_1-k_2) & \quad
\{\rho(k_1),\rho(k_2)\} &=\{\phi(k_1),\phi(k_2)\}=0,    \\
\{\rho_m,\phi_n\} &= \delta_{mn}, \label{kdv-eq57} &\quad
\{\rho_m,\rho_n\} &=\{\phi_m,\phi_n\}=0.
\end{align}
($k_{1}>0, k_{2}>0$) with respect to the Poisson bracket:
\[
\{A,B\}=\Bigglb \frac{\delta A}{\delta u},\frac{\delta B}{\delta
u}\Biggrb .
\]
One can verify that the scattering data determines the canonical
variables and vice-versa. The symplectic basis consists of the
eigenfunctions of the operator $\Lambda={1\over 2}(L_++L_-)$:
\begin{eqnarray}
\Lambda \mathcal{P}(x,k)&=&k^2 \mathcal{P}(x,k), \\
\Lambda \mathcal{Q}(x,k)&=&k^2 \mathcal{Q}(x,k), \\
\Lambda P_n(x)&=& k_n^2 P_n(x), \\
\Lambda Q_n(x)&=& k_n^2 Q_n(x).
\end{eqnarray}
Again, for variations due to the time-evolution of $u$, $\delta
u(x,t)=u_t \delta t + Q((\delta t)^2)$, etc. the equations of the
KdV hierarchy \begin{eqnarray}u_t+\partial_x
\Omega(\Lambda)u(x,t)=0, \label{eq:KdVHL}
\end{eqnarray} with (\ref{kdv-eq:q-PQ}) and (\ref{eq:kdv-du-PQ}) are equivalent to a
system of trivial linear ordinary differential equations for the
canonical variables (which can be considered as scattering data):
\begin{equation}\label{AA}\begin{aligned}
\phi_t&= 4i\beta(k)\Omega(k^2),  &\qquad \rho_t(k)&=0,  \\
\phi_{n,t}&= 4ik_n\Omega(k_n^2), &\qquad \rho_{n,t}&=0.
\end{aligned}\end{equation}

\section{Perturbations to the equations of the KdV hierarchy}\label{sec:7}

\subsection{General perturbations}

Let us consider a general perturbation $W_x[u]$ to an equation
from the KdV hierarchy:
\begin{eqnarray}u_t+\partial_x
\Omega(\Lambda)u(x,t)=W_x[u]. \label{eq:KdVH-pert}
\end{eqnarray}
The function $W_x[u]$ is assumed to belong to the class of
admissible potentials for the associated spectral problem
(\ref{SPKdV}) (Schwartz class functions, in our case).

The expansion of the perturbation over the symplectic basis is:
\begin{eqnarray}\label{eq:kdv-W-PQ}
    W_x[u]&=& \frac{1}{2\pi}\int_0^\infty \frac{{\rm d}k}{\beta(k)}\, \left( \mathcal{P}_{x}(x,k)
    \phi_W(k) - \mathcal{Q}_{x}(x,k)\rho_W(k)\right)  \\
&+&\phi_0 \mathcal{P}_x(x,0)    - \sum_{n=1}^N    \left(
P_{n,x}(x)\phi_{n,W} - Q_{n,x}(x)\rho_{n,W}\right).\nonumber
\end{eqnarray}
The substitution of the above expansion (\ref{eq:kdv-W-PQ}) in
(\ref{eq:KdVH-pert}) together with (\ref{kdv-eq:q-PQ}) and
(\ref{eq:kdv-du-PQ}) leads to a modification of the time evolution
(\ref{AA})  of the scattering data as follows:
\begin{eqnarray}
\phi_t&=&4i\beta(k)\Omega(k^2)+\phi_W(k,t;\rho(k,t), \phi(k,t), \rho_n(t),\phi_n(t))+ \phi_0 \delta(k),\label{AAW 1}  \\
\rho_t(k)&=&\rho_W(k,t;\rho(k,t), \phi(k,t), \rho_n(t),\phi_n(t)), \label{AAW 2}\\
\phi_{n,t}&=&4ik_n\Omega(k_n^2)+\phi_{n,W}(k,t;\rho(k,t), \phi(k,t), \rho_n(t),\phi_n(t)), \label{AAW 3} \\
\rho_{n,t}&=&\rho_{n,W}(k,t;\rho(k,t), \phi(k,t),
\rho_n(t),\phi_n(t)).\label{AAW 4}
\end{eqnarray}
Since $W=W[u]$ and $u$ depend on the scattering data, we observe
that the expansion coefficients of the perturbation ($\phi_W(k) =
\biglb \mathcal{Q}(y,k),W(y)\bigrb $ etc.) also depend on the
scattering data. Thus for generic $W$ the new dynamical system
(\ref{AAW 1}) -- (\ref{AAW 4}) for the scattering data can be
extremely complicated and non-integrable in general. This reflects
the obvious fact that the perturbed integrable equations are, in
general, not integrable.

Below we consider several important particular cases.

\subsection{Self-consistent sources -- integrable case}

It is widely spread  that  KdV with self-consistent sources is a
special case of perturbed KdV with
\begin{equation}\label{eq:scs-1}\begin{split}
   W_x[u]=  \sum_{n=1}^N    c_n \mathcal{P}_{n,x}(x)  + c_0 \mathcal{P}_{x}(x,0),
\end{split}\end{equation}
where $c_n$ are some constants and the constant $c_0$ is usually
set to 0. The last term in (\ref{eq:scs-1}) affects only the
end-point of the continuous spectrum of $L$. Such choice of the
perturbation substantially simplifies the evolution of the
scattering data:
\begin{equation}\label{eq:sde-1}\begin{aligned}
\phi_t&=4i\beta(k)\Omega(k^2) -c_0 \delta(k), &\qquad  \rho_t(k)&=0 \\
\phi_{n,t}&= 4ik_n\Omega(k_n^2)- c_n , &\qquad \rho_{n,t}&= 0.
\end{aligned}\end{equation}
It is obvious that perturbations (\ref{eq:scs-1}) preserve
integrability. Indeed, from eqs. (\ref{eq:sde-1}) one finds that
such perturbations are isospectral, since $k_{n,t}=0$. The only
effect of the perturbation (\ref{eq:scs-1}) consists in changing
the time-dependence of the angle variables, related to the
discrete spectrum of $L$.

\begin{example}\label{exa:1}

Let $c_0 \neq 0$ and all other $c_n =0$. Then $W_x[u] =c_0
P(x,0)$. Since $\Lambda P(x,0) =0$ (see eq. (\ref{kdv-eq53})) we
easily conclude, that
\begin{equation}\label{eq:pKdv6}\begin{aligned}
\Lambda (v_t +v_{xxx} -3(v^2)_x - c_0 \mathcal{P}(x,0)) = \Lambda
(v_t +v_{xxx} -3(v^2)_x)=0,
\end{aligned}\end{equation}
i.e such perturbed KdV is equivalent to the KdV6 equation. Thus
from the above considerations we confirm the result in
\cite{bak08}, that KdV6  is integrable.
\end{example}

\subsection{Self-consistent sources -- non-integrable case}

Here we consider a more general type of self-consistent sources
given by:
\begin{equation}\label{eq:scs-2}\begin{split}
   W_x[u]=  \sum_{n=1}^N    ( c_n \mathcal{P}_{n,x}(x)  + \tilde{c}_n \mathcal{Q}_{n,x}(x)) + c_0 \mathcal{P}_{x}(x,0),
\end{split}\end{equation}
Again we consider $c_n$, $\tilde{c}_n$ and $c_0$ as constants.
Note, that $\mathcal{Q}_{x}(x,k)$ is an odd function of $k$ and
therefore  $\mathcal{Q}_{x}(x,0)=0$. The evolution of the
scattering data under such perturbation is:
\begin{equation}\label{eq:sde-2}\begin{aligned}
\phi_t&=4i\beta(k)\Omega(k^2) -c_0 \delta(k), &\qquad  \rho_t(k)&=0 \\
\phi_{n,t}&= 4ik_n\Omega(k_n^2)- c_n , &\qquad \rho_{n,t}&=
\tilde{c}_n.
\end{aligned}\end{equation}
i.e both the action variables $\rho_n$ and the discrete
eigenvalues become time-dependent. Such perturbation is not
isospectral and obviously violates integrability. In addition the
angle variables become nonlinear functions of $t$.

Let us investigate the integrability of the following
equation:\begin{eqnarray}\label{5.01} \Lambda^*(u_t+\partial_x
\Omega(\Lambda)u(x,t))=0,
\end{eqnarray}
where the star is a notation for a Hermitian conjugation. KdV6 in
(\ref{KdV6-1}) is a particular case of this equation with
$\Omega(\Lambda)=-4\Lambda$. In order to simplify our further
analysis, instead of the equation (\ref{5.01}) we study the
following one:
\begin{eqnarray}\label{5.02} (\Lambda^*-\lambda_1)(u_t+\partial_x
\Omega(\Lambda)u(x,t))=0,
\end{eqnarray}
where $\lambda_1$ is a constant. The corresponding analogue for
KdV6 is
\begin{multline*}
v_{6x}+ v_{txxx}-2v_t v_{xx}-
4v_xv_{xt}- 10 v_xv_{4x}-20v_{xx}v_{xxx}+30 v_x^2v_{xx}  \\
+ 4\lambda_1(v_{xt}+v_{xxxx}-6v_xv_{xx})=0.
\end{multline*}
Due to the identity $\partial \Lambda=\Lambda^* \partial$ we can
represent (\ref{5.02}) in the form \begin{eqnarray}\label{5.03}
\partial(\Lambda-\lambda_1)((\partial_x^{-1}u_t)+ \Omega(\Lambda)u(x,t))=0.
\end{eqnarray}
Since the operator $\partial $ does not have a kernel when $u$ is
Schwartz class, (\ref{5.03}) is equivalent to
\begin{equation}\label{5.12} u_t+\partial_x \Omega(\Lambda)u(x,t) =
\left \{\begin{aligned}
 & \left(c_1P_1(x,t) +\tilde{c}_1Q_1(x,t)\right)_x &    &\mbox{for} &  \lambda_1 &=k_1^2<0,\\
& \left(c_1 \mathcal{P}(x,k_1,t)
+\tilde{c}_1\mathcal{Q}(x,k_1,t)\right)_x &  &\mbox{for} &
\lambda_1&=k_1^2>0,
\end{aligned} \right.
\end{equation}
where
\begin{eqnarray*}
c_{1}&=&c_{1}(t,k_1;\rho(k_1,t), \phi(k_1,t),
\rho_n(t),\phi_n(t)),\\
\tilde{c}_{1}&=&\tilde{c}_{1}(t,k_1;\rho(k_1,t), \phi(k_1,t),
\rho_n(t),\phi_n(t))
\end{eqnarray*}
are $x$-independent functions, but the important observation is
that the time-depen\-den\-ce could be implicit through the
scattering data of the potential $u(x,t)$. Equation (\ref{5.12})
is a perturbed equation from the KdV hierarchy. The perturbation
in the right-hand side of (\ref{5.12}) is in the eigenspace of the
recursion operator corresponding to the eigenvalue $\lambda_1$,
i.e. it is given by  'squared' eigenfunctions of the spectral
problem (\ref{SPKdV}) at $\lambda_1$.

If $\lambda_1<0$ is a discrete eigenvalue, the corresponding
dynamical system (\ref{AAW 1}) -- (\ref{AAW 4}) for the scattering
data is
\begin{eqnarray}
\phi_t&=&4i\beta(k)\Omega(k^2),\label{SC 1} \\
\rho_t(k)&=&0, \label{SC 2}\\
\phi_{n,t}&=&4ik_n\Omega(k_n^2)-c_1(t,k_1;\rho(k_1,t),
\phi(k_1,t),
\rho_n(t),\phi_n(t))\delta_{n,1}, \label{SC 3} \\
\rho_{n,t}&=&\tilde{c}_1(t,k_1;\rho(k_1,t), \phi(k_1,t),
\rho_n(t),\phi_n(t))\delta_{n,1}.\label{SC 4}
\end{eqnarray}
If $\lambda_1>0$ is a continuous spectrum eigenvalue, the
dynamical system (\ref{AAW 1}) -- (\ref{AAW 4}) has the form
\begin{eqnarray}
\phi_t&=&4i\beta(k)\Omega(k^2) \nonumber \\ &+& 2\pi \beta(k)
c_1(t,k_1;\rho(k_1,t), \phi(k_1,t),
\rho_n(t),\phi_n(t))\delta(k-k_1),\label{SC 1a} \\
\rho_t(k)&=&-2\pi \beta(k)\tilde{c}_1(t,k_1;\rho(k_1,t),
\phi(k_1,t),
\rho_n(t),\phi_n(t))\delta(k-k_1), \label{SC 2a}\\
\phi_{n,t}&=&4ik_n\Omega(k_n^2), \label{SC 3a} \\
\rho_{n,t}&=&0.\label{SC 4a}
\end{eqnarray}
It is clear, that dynamical systems like (\ref{SC 1}) -- (\ref{SC
4}) or (\ref{SC 1a}) -- (\ref{SC 4a}) can not be integrable for a
general functional dependence of $c_{1}$ and $\tilde{c}_1$ on the
scattering data. Thus the generic perturbed equations from KdV
Hierarchy(\ref{5.01}),  are not {\it completely} integrable. In
other words, there are solutions, which can not be obtained via
the Inverse Scattering Method, since the aforementioned dynamical
systems for the scattering data are not always integrable.

\begin{example}\label{exa:3}
Let us assume that $\lambda_1=-\kappa_1^2<0$ is a discrete
eigenvalue of $L$, and that $c_1$ and $\tilde{c}_1$ in eq.
(\ref{5.12}) are constants. Then the corresponding perturbed KdV
will be equivalent to the following equations for the scattering
data of $L$:
\begin{equation}\label{eq:scs-3}\begin{aligned}
\phi_t&= 4i\beta(k)\Omega(k^2),  &\qquad  \rho_t(k)&=0,  \\
\phi_{n,t}&= 4ik_n\Omega(k_n^2)-c_1 \delta_{n,1}, &\qquad
\rho_{n,t} &= \tilde{c}_1 \delta_{n,1}.
\end{aligned}\end{equation}
From the last of  these equations we find
\begin{equation}\label{eq:k1t}\begin{split}
\kappa_1^2(t) = \tilde{c}_1 t + \kappa_1^2(0),
\end{split}\end{equation}
i.e the eigenvalue $\kappa_1(t)$ explicitly depends on $t$.
Clearly (\ref{eq:k1t}) makes sense for those $t$ for which
$\tilde{c}_1 t + \kappa_1^2(0)>0$. Then (i) if $\tilde{c}_1 t +
\kappa_1^2(0)<0$ the eigenvalue overlaps with the continuous
spectrum of $L$ and the solution develops singularities; (ii) if
for some $t$ $\kappa_1$ crosses another discrete eigenvalue the
spectrum cease to be simple and the solution is not in the assumed
Schwartz class, i.e. again may develop singularities.

\end{example}

\begin{example}\label{exa:4}
Let us assume that $\lambda_1=-\kappa_1^2<0$ is the discrete
eigenvalue of $L$, and that $c_1$ is constant and $\tilde{c}_1$ is
a function of time, for example:
\begin{equation}\label{eq:k2t}\begin{split}
\tilde{c}_1(t) = \epsilon_0 \kappa_1^2(0) \omega \cos (\omega t)
\end{split}\end{equation}
where $\epsilon_0 \ll 1$ and $\omega_0$ are constants. Then the
eigenvalue $\kappa_1^2$ depends explicitly on $t$ according to
\begin{equation}\label{eq:scs-3exa}\begin{aligned}
\kappa_1^2(t) = \kappa_1^2(0) (1 +\epsilon_0 \sin (\omega t)).
\end{aligned}\end{equation}
If $\epsilon_0$ is chosen small enough the corresponding
eigenvalue will not overlap with the continuous spectrum, nor with
other discrete eigenvalues. Otherwise - as above and again we may
have singularities.

\end{example}

\section{ Examples of exact solutions}\label{sec:6}

As we have seen, KdV6 in general is not completely integrable
system. However, there are many solutions of (\ref{5.01}) or
(\ref{5.03}) that can be written explicitly (apart from the KdV
solutions, which are an obvious subclass of solutions).

One such example is when $c_1$ and $\tilde{c}_1$ depend only on
$t$. Then the system (\ref{SC 1}) -- (\ref{SC 4}) is integrable,
with new canonical variables \begin{eqnarray}
\tilde{\phi}_n&=&\phi_n(0)+4ik_n\Omega(k_n^2)t-\delta_{n,1}\int_{0}^{t}c_1(\tau)d\tau,
\nonumber \\
\tilde{\rho}_{n}&=&\rho_n(0)+\delta_{n,1}\int_{0}^{t}\tilde{c}_1(\tau)d\tau.
\nonumber
\end{eqnarray}
Since $c_{1}(t)$ and $\tilde{c}_1$ do not depend on the scattering
data and therefore on $u(x,t)$, the Poisson brackets between
$\tilde{\phi}_n$ and $\tilde{\rho}_{n}$ are canonical.

This is most often the situation that many authors assume about
KdV6 (i.e. $c_{1}=c_{1}(t)$ and $\tilde{c}_1=\tilde{c}_1(t)$
simply given functions of $t$) when discuss its integrability.

There are, however other exactly solvable cases. Indeed, if
\begin{eqnarray} c_1&=&\sum_{k=1}^{N}
(\alpha_{1k}\phi_k+\beta_{1k}\rho_k),
\nonumber \\
\tilde{c}_1&=&\sum_{k=1}^{N} (\alpha_{2k}\phi_k+\beta_{2k}\rho_k)
\nonumber
\end{eqnarray}
are linear combinations of the scattering data with some constants
$\alpha_{p,q}$, $\beta_{{p,q}}$ the system (\ref{SC 1}) --
(\ref{SC 4}) becomes a system of linear ordinary ODEs  and
therefore is integrable.

To illustrate the effect of non-holonomic deformations on the
soliton solutions, let us consider the 1-soliton solutions
corresponding to the specific choice of perturbation functions
$c_1$ and $\tilde{c}_1$ considered in examples 2 and 3 in the
previous  Section. The one-soliton solution for the KdV equation
(corresponding to a discrete eigenvalue $\lambda_1=-\kappa_1^2$)
is given by \cite{ZMNP}:
\begin{eqnarray}\label{eq:1s}
u_{\rm 1s}(x,t)=-{2\kappa_1^2\over {\rm ch}^2\,
\kappa_1\left(x-4\kappa_1^2t-{\phi_1\over \kappa_1}\right)},
\end{eqnarray}
where $\phi_1$ is the angle variable that corresponds to the
prescribed eigenvalue $\lambda_1=-\kappa_1^2$. Here, also, the
quantity $v=4\kappa_1^2$ can be interpreted as a velocity of the
soliton, $\phi_1$ -- as a position of soliton's mass-center at
$t=0$ (it is often called a soliton phase).

For the special choice of constants in {\it Example 2} one can
write down the corresponding perturbed 1-soliton solutionin the
following form:
\begin{eqnarray}\label{eq:1s-p1}
\tilde{u}_{\rm 1s}(x,t)=-{2(\tilde{c}_1t+\kappa_1(0)^2)\over {\rm
ch}^2\,
\left\{\sqrt{\tilde{c}_1t+\kappa_1(0)^2}\left[x-\left(4\kappa_1^2-{\tilde{c}_1\over
\kappa_1}\right)t-{\phi_1\over \kappa_1}\right]\right\}}.
\end{eqnarray}
Here the correction in the velocity of the soliton
$\tilde{v}=v+{\tilde{c}_1\over \kappa_1}$ is due to shift in the
angle variable $\phi_1$.

For the case of  {\it Example 3} the corresponding perturbed
1-soliton solution reads:
\begin{eqnarray}\label{eq:1s-p2}
\tilde{u}_{\rm 1s}(x,t)={2\kappa_1(0)^2(1+\epsilon_0\sin \omega
t)\over {\rm ch}^2\, \left\{\kappa_1(0)\sqrt{1+\epsilon_0\sin
\omega t}\left[x+4\kappa_1^2(0) (1 +\epsilon_0 \sin \omega
t)t-{\phi_1\over \kappa_1}\right]\right\}}.
\end{eqnarray}
Here the soliton velocity depends on time, which is typical for
the so-called boomerons \cite{cd82}. Note however, that the
boomeron type behavior is characteristic for multi-component KdV
equations, while here we get it as an effect  of the perturbation.
Moreover, the time dependence of the velocity is controlled by the
time-depending coefficient in the perturbation $\tilde{c}_1(t)$.

There are of course many other nontrivial integrable examples that
one can construct by various choices of the functions $c_{1}$ and
$\tilde{c}_1$ -- when (\ref{SC 1}) -- (\ref{SC 4}) is an
integrable system of ODEs. In general, however this system is not
integrable and therefore there exist solutions that can not be
constructed by the means of IST.

\section{Discussions and Conclusions}\label{sec:8}

Here we have used the expansion over the eigenfunctions of the
recursion operator for the KdV hierarchy for studying nonholonomic
deformations of the corresponding NLEE from the hierarchy. We have
shown, that in the case of self-consistent sources, the
corresponding perturbed NLEE is integrable, but not completely
integrable.

The perturbation results for the Zakharov-Shabat (ZS) type
spectral problems have been obtained firstly in \cite{K76} and for
KdV in \cite{KM77}. As it has been explained, the perturbation
theory is based on the completeness relations for the squared
eigenfunctions. For the Sturm-Liouville spectral problem such
relations apparently have been studied as early as in 1946
\cite{B46} and then by other authors, e.g. \cite{KKh81,IKK94}. The
completeness relation for the eigenfunctions of the ZS spectral
problem is derived in \cite{K76a} and generalisations are studied
further in \cite{G86,GI92,GeHr1,g1994,GY94,TV08}, see also
\cite{gvy2008} (and the references therein).

The approach presented in this article can be applied also to the
study of inhomogeneous versions of NLEE, related to other linear
spectral problems, e.g. the Camassa-Holm equation
\cite{CGI,CGI07},  various difference and matrix generalizations
of KdV-like  and Zakharov - Shabat spectral problems, various
non-Hamiltonian systems \cite{wang09}, etc.

\section*{Acknowledgments} The authors have the pleasure to thank Prof. Adrian Constantin,
Prof. Nikolay Kostov, Prof. Anjan Kundu, Prof. Alexander Mikhailov
and Dr. Jing Ping Wang for numerous useful discussions. Part of
this work was done in the Erwin Schr\"odinger International
Institute for Mathematical Physics (Vienna) during the authors
participation in the programme 'Nonlinear Water Waves', April -
June 2011. This material is based upon works supported by the
Science Foundation Ireland (SFI), under Grant No. 09/RFP/MTH2144.


\end{document}